\documentclass[preprint,tightenlines,eqsecnum,floats,aps,amsmath,amssymb,nofootinbib,prd,showpacs,superscriptaddress]{revtex4}  
\usepackage{amsmath,amssymb,amsfonts}
\usepackage{graphicx}
\usepackage{enumerate}
\usepackage[colorlinks=true,linkcolor=blue,anchorcolor=blue,citecolor=blue,urlcolor=blue]{hyperref}
\usepackage[sort&compress]{natbib}
\usepackage{url}
\usepackage[T1]{fontenc}
\usepackage{mathptmx}
\usepackage{float}
\usepackage{booktabs}
\usepackage{CJK}
\usepackage{epstopdf}
\usepackage{tikz}
\usetikzlibrary{positioning, shapes.geometric, arrows}

\begin{document}
\title{From masses and radii of neutron stars to EOS of nuclear matter through neural  network}

\author{Zehan Wu}
\affiliation{School of Physics and Optoelectronics, South China University of
Technology, Guangzhou 510641, P.R. China}
\author{ Dehua Wen\footnote{Corresponding author. wendehua@scut.edu.cn}}
\affiliation{School of Physics and Optoelectronics, South China University of
Technology, Guangzhou 510641, P.R. China}
\date{\today}

\begin{abstract}
The equation of state (EOS) of dense nuclear matter is a key factor to determine the internal structure and properties of neutron stars. However, the EOS of high-density nuclear matter has great uncertainty mainly because the terrestrial nuclear experiments cannot reproduce matter as dense as that in the inner core of a neutron star. Fortunately, continuous  improvements in astronomical observations of neutron stars  provide the opportunity to inversely constrain the EOS of high-density nuclear matter. A number of methods have been proposed to implement this inverse constraint, such as the Bayesian analysis algorithm, the Lindblom's approach, and so on. Neural network algorithm is an effective new method developed in recent years.
By employing a set of isospin-dependent parametric EOSs as the training sample of neural network algorithm, we set up an effective way to  reconstruct the EOS with relative accuracy through a few mass-radius data. Based on the obtained neural network algorithms and according to the NICER  observations on masses and radii of neutron stars with assumed precision, we get the inversely constrained EOS and further calculate the corresponding macroscopic properties of the neutron star. The results are  basically consistent with the constraint on EOS from the Huth $et~ al.$ based on Bayesian analysis. Moreover, the results show that even though the neural network algorithm was obtained by using the finite parameterized EOS as the training set, it is valid for any rational parameter combination of the parameterized EOS model.

\end{abstract}

\pacs{97.60.Jd; 04.40.Dg; 04.30.-w; 95.30.Sf}

\maketitle

\section{Introduction}
Understanding the nature of neutron stars and determining the equation of state (EOS) of high-density asymmetric nuclear matter are both long-standing problems in nuclear physics and astrophysics \cite{Chen1, Li25, Li26,Li2019EPJA}. The EOS of asymmetric nuclear matter is a key input for probing the structure and properties of neutron stars. However, the theoretical predictions on the EOS at supra-saturation densities diverge broadly. Except possible phase transitions,  density dependence of the
symmetry energy is the most uncertain part of the EOS, especially at supra-saturation densities \cite{Dieperink03PRC,Li26,Li2019EPJA}.
Although significant progresses have  been made in probing the symmetry energy in  terrestrial nuclear laboratories \cite{Russotto2,LeFevre3,Adhikari4}, the inability to reproduce the authentic high-density neutron-rich nuclear matter restricts the full and accurate understanding of symmetry energy and EOS at supra-saturation densities.

Fortunately, many groundbreaking observational discoveries from  scientific facilities in recent years have led to a new upsurge in understanding the  neutron stars and  EOS of high-density asymmetric nuclear matter. For example, the mass-radius measurement on neutron stars from the Neutron Star Interior Composition Explorer (NICER) and the gravitational radiation detection of binary neutron star mergers from LIGO and Virgo have taken the understanding of neutron stars and EOS of the asymmetric nuclear matter to a new level \cite{Miller5,Riley6,Pang7,Abbott8}. In the near future, the Large Observatory for X-ray Timing (LOFT), the Advanced Telescope for High Energy Astrophysics (ATHENA), the Square Kilometre Array (SKA) and Einstein Telescope (ET) will also bring more observations of neutron stars when they become operational  \cite{Feroci40, Nandra41, Punturo42, Dewdney43, Rezzolla44}. It is believed that they will provide powerful insights in understanding the neutron stars and the EOS.

If we regard obtaining the properties of neutron stars from the EOS of nuclear matter as forward method, which normally can be numerically calculated according to the EOS and TOV equations for static spherically symmetrical stars. In combination with the inverse derivation from the mass-radius to EOS given in Ref. \cite{Lindblom9} using enthalpy, there is a one-to-one mapping between EOS and the mass-radius relation of neutron stars. That is, if a set of accurate mass-radius observations of neutron stars is obtained, the EOS of nuclear matter can be reversely achieved. But how can we reversely obtain the EOS from the properties of neutron stars?  At present, there is no reliable and widely accepted  method. The purpose of this work is to try to find an effective way reversely mapping the EOS of nuclear matter through the mass-radius relations of neutron stars.

Lindblom had pioneered the inverse TOV mapping to obtain the pressure-density sequence P($\rho$) of the EOS from the accurate mass-radius sequence of a neutron star \cite{Lindblom9}. The reality is that the accurate observation data of the mass-radius of the neutron star is still very rare. As more advanced algorithms are introduced, a small amount of observation data can provide effective constraint on the EOS. For example, Ref. \cite{XuJPRC21} implemented constraints on EOS-related parameters within the framework of Bayesian analysis. Bayesian analysis is an algorithm for updating information by correcting existing information based on observed events. The use of this algorithm therefore requires ensuring both the correctness of the prior probabilities and the reasonableness of the observed events. This year, Huth $et ~al.$ also used this algorithm to achieve constraints on the EOS by combining data from heavy-ion collisions (HIC) with results from multi-messenger astronomical observations \cite{Huth11}. Furthermore, a substantial amount of Bayesian analysis work has been contributed to neutron star research \cite{YX12,Silva13,Xie14}.

Neural networks (NN) is a branch of machine learning (ML) and a computing system inspired by biological neural networks \cite{Amari17}. It has the advantages of Bayesian analysis algorithm and polynomial fitting algorithm. Only a few less precise observations are needed to obtain the effective constraint through NN, which can also be used to calculate other properties of neutron stars. In the 1990s, NN was pioneered by Clark and collaborators for research related to nuclear physics \cite{Gazula92,Gazula93,Gazula95,Gazula99}. This algorithm is now widely used in theoretical prediction work to improve the accuracy of nuclear mass and nuclear charge radii measurements \cite{ZhaoNPA22,Niu18,Utama16PRC,Utama16JPG}. In the field of nuclear astrophysics, this algorithms have also been introduced into neutron star research. For example, the inverse TOV mapping of NN was constructed by using the EOS with sound velocity segmentation in Ref. \cite{Fujimoto16},  the reconstructed EOS was implemented with NN at 1 to 7 times the saturated nuclear density $\rho _0\approx 0.16\ \mathrm{fm}^{-3}$ in Ref. \cite{Soma22AX},  Krastev constructed NNs from mass-radius and mass-tidal deformations to symmetry energy \cite{Krastev15}, and NN from nuclear parameters to neutron star properties was constructed in Ref. \cite{Thete22AX}. Additionally, the mapping from X-ray spectra to EOS parameters was achieved by using NN in Ref. \cite{Farrell22AX}.

Although much of the NN work has given exciting results, many questions remain unsolved. For example, the plausibility of the symmetry energy parameters of the EOS after noise addition, the implementation of the NN prediction function and its general applicability, and the validation of the effectiveness. In this work, we will  construct a new NN based on the parametric EOS, which can be efficiently implemented for any combination of parameters satisfying terrestrial experiments and multi-messenger astronomical observations with a small number of mass-radius relations to inverse TOV mapping.

This paper is organized as follows. In Sec. \ref{s1} we briefly review the isospin-dependent parametric EOS and the basic formula for calculating the properties of neutron stars.   The NN implementation of the inverse TOV mapping is presented in Sec. \ref{s2}. NN combined with NICER observational constraints  are presented in Sec. \ref{s3}.  The summary and outlook are in Sec. \ref{s4}.

Unless otherwise stated, in the formulae we use the gravitational units ($G = c = 1$).

\section{Isospin-dependent parametric EOS and the fundamental formula of neutron stars} \label{s1}

\subsection{Isospin-dependent parametric EOS} \label{ISOEOS}
Implementing a NN for inverse TOV mapping requires a large number of the EOS as training samples. Moreover, these EOS also need to meet the range of nuclear parameters as much as possible, such as $L$, $K_{\mathrm{sym}}$, etc. The isospin-dependent parametric EOS is able to satisfy both the need to generate a large number of EOS and the rationalization of nuclear parameters.

For asymmetric nuclear matter, the energy per nucleon can be expanded by isospin asymmetry \cite{Li25, Li26}
\begin{equation}\label{2.1}
E\left( \rho ,\delta \right) \approx E_0\left( \rho \right) +E_{\mathrm{sym}}\left( \rho \right) \delta ^2,
\end{equation}
where $\delta =\left( \rho _\mathrm{n}-\rho _\mathrm{p} \right) /\left( \rho _\mathrm{n}+\rho _\mathrm{p} \right) $ is the isospin asymmetry. The first term on the right side $E_0\left( \rho \right)$ is usually referred to as the energy of symmetric nuclear matter, and the second term $E_{\mathrm{sym}}\left( \rho \right)$ is referred to as the symmetry energy, which has the physical meaning of the energy difference per nucleon between asymmetric nuclear matter and pure neutron matter.

Usually, the two terms on the right side in Eq. (\ref{2.1}) can be expanded at the saturation density $\rho_0$ \cite{Zhang18APJ, Xie14}
\begin{gather*}
E_0\left( \rho \right) =E_0\left( \rho _0 \right) +\frac{K_0}{2}\left( \frac{\rho -\rho _0}{3\rho _0} \right) ^2+\frac{J_0}{6}\left( \frac{\rho -\rho _0}{3\rho _0} \right) ^3,\tag{2.2}
\\
E_{\mathrm{sym}}\left( \rho \right) =E_{\mathrm{sym}}\left( \rho _0 \right) +L\left( \frac{\rho -\rho _0}{3\rho _0} \right) +\frac{K_{\mathrm{sym}}}{2}\left( \frac{\rho -\rho _0}{3\rho _0} \right) ^2+\frac{J_{\mathrm{sym}}}{6}\left( \frac{\rho -\rho _0}{3\rho _0} \right) ^3.\tag{2.3}
\end{gather*}

Constrained by terrestrial nuclear experiments, the most probable values of the parameters in the above equations are as follows: $L=58.7\pm 28.1\ \mathrm{MeV}$, $-400\leqslant K_{\mathrm{sym}}\leqslant 100\ \mathrm{MeV}$, $-200\leqslant J_{\mathrm{sym}}\leqslant 800\ \mathrm{MeV}$, $-300\leqslant J_0\leqslant 400\ \mathrm{MeV}$, $K_0=240\pm 20\ \mathrm{MeV}$, $E_{\mathrm{sym}}\left( \rho _0 \right) = 31.7\pm 3.2\ \mathrm{MeV}$ \cite{Shlomo33, Piekarewicz34, Zhang35, Oertel36, Li37}. It is worth pointing out that in recent years, PREX-\uppercase\expandafter{\romannumeral2} has given higher $L$ values ($L=106\pm 37\ \mathrm{MeV}$ \cite{Adhikari4}). This result is also considered later in this work.

The pressure of the system can be calculated numerically from
\begin{equation}
P\left( \rho ,\delta \right) =\rho ^2\frac{d\left( E/\rho \right)}{d\rho}.\tag{2.4}
\end{equation}

To construct the EOS in the whole density range, the NV EOS model \cite{Negele38} and BPS EOS model \cite{Baym39} are used for the inner crust and the outer crust, respectively.

\subsection{Fundamental formula of statically spherically symmetric neutron stars}
Neutron stars in this work are considered to be isolated, non-rotating, and statically spherically symmetric. The Tolman-Oppenheimer-Volkoff (TOV) equations consist of the equation of hydrostatic equilibrium \cite{Oppenheimer19,Tolman20}
\begin{equation}\label{2.4}
\frac{dP}{dr}=-\frac{\left[ m\left( r \right) +4\pi r^3P\left( r \right) \right] \left[ \rho \left( r \right) +P\left( r \right) \right]}{r\left[ r-2m\left( r \right) \right]}, \tag{2.5}
\end{equation}

and

\begin{equation}
\frac{dm\left( r \right)}{dr}=4\pi r^2\rho \left( r \right). \tag{2.6}
\end{equation}

The external boundary condition of the star is as follows: $\rho=p=0$. Given the center density of the neutron star, it can be solved layer by layer from the interior to the exterior of the star using the method of solving high-precision initial value problems.

Tidal deformation of neutron star is an EOS-dependent property which can be measured through the gravitational wave events such as GW170817 \cite{Abbott18PRL}. To linear order, the tidal deformation $\varLambda$ is defined as \cite{Flanagan22}
\begin{equation}
Q_{ij}=-\varLambda \varepsilon _{ij}, \tag{2.7}
\end{equation}
where $Q_{\mathrm{ij}}$ and $\varepsilon _{\mathrm{ij}}$ represent the induced quadrupole moment and the static tidal field, respectively. The second tidal Love number $k_2$ is defined as
\begin{equation}
k_2=\frac{3}{2}\varLambda R^{-5}, \tag{2.8}
\end{equation}
which can be solved together with the TOV equation \cite{Chamel23, Chamel24}.

\section{Neural network implementation of inverse TOV mapping} \label{s2}
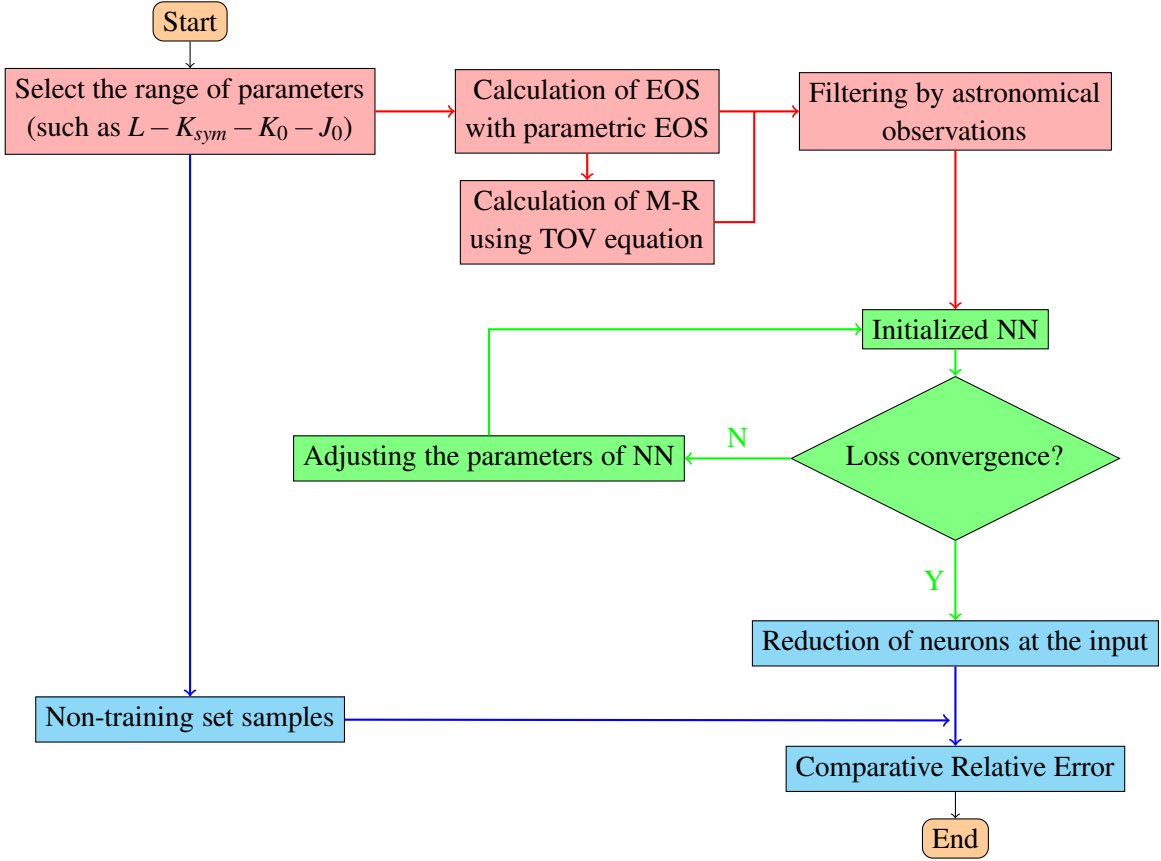
\begin{figure}[htp]
    \centering
    \begin{tikzpicture}[node distance=10pt]
      \node[draw, align=center, rounded corners, fill=orange!40]   (start)   {Start};
      \node[draw, align=center, below=of start,fill=red!30]    (step 1)  {Select the range of parameters \\(such as $L-K_{sym}-K_{0}-J_{0}$)};
      \node[draw, align=center, right=30pt of step 1,fill=red!30]     (step 2)  {Calculation of EOS \\with parametric EOS};
      \node[draw, align=center, below=of step 2,fill=red!30]     (step 3) {Calculation of M-R \\using TOV equation};
      \node[draw, align=center, right=30pt of step 2,fill=red!30]   (step 4)    {Filtering by astronomical \\observations};
      \node[draw, align=center, below=60pt of step 4,fill=green!50]    (step 5) {Initialized NN};
      \node[draw, diamond, aspect=2, below=of step 5,fill=green!50]     (choice)  {Loss convergence?};
      \node[draw, align=center, left=40pt of choice,fill=green!50]  (step 6)  {Adjusting the parameters of NN};
      \node[draw, align=center, below=30pt of choice,fill=cyan!40]   (step 7)    {Reduction of neurons at the input};
      \node[draw, align=center, below=30pt of step 7,fill=cyan!40]   (step 8)  {Comparative Relative Error};
      \node[draw, rounded corners, below=of step 8, fill=orange!40]  (end)     {End};
      \node[draw, align=center, below=205pt of step 1,fill=cyan!40]   (step 9)    {Non-training set samples};
      
      \draw[->] (start)  -- (step 1);
      \draw[color=red,thick,->] (step 1) -- (step 2);
      \draw[color=red,thick,->] (step 2) -- (step 3);
      \draw[color=red,thick] (step 3) -- (7.5,-2.67) -- (7.5,-1.2);
      \draw[color=red,thick,->] (step 2) -- (step 4);
      \draw[color=red,thick,->] (step 4) -- (step 5);
      \draw[color=green,thick,->] (step 5) -- (choice);
      \draw[color=green,thick,->] (choice) -- node[above] {N}  (step 6);
      \draw[color=green,thick,->] (step 6) -- (step 6|-step 5) -> (step 5);
      \draw[color=green,thick,->] (choice) -- node[left]  {Y} (step 7);
      \draw[color=blue,thick,->] (step 7) -- (step 8);
      \draw[->] (step 8) -- (end);
      \draw[color=blue,thick,->] (step 1) -- (step 9);
      \draw[color=blue,thick,->] (step 9) -- (10.1,-9.3);

    \end{tikzpicture}
    \caption{Flowchart of inverse TOV mapping neural network implementation}
    \label{Flowchat}
 \end{figure}

To clearly demonstrate the process of implementing an NN for inverse TOV mapping, we draw the flow chart as shown in Fig. \ref{Flowchat}. The first step requires the provision of a relatively complete training set. After determining the range of symmetry energy parameters, the massive EOS (output of NN) is generated by the method in Section \ref{ISOEOS}, while the corresponding mass-radius points (input of NN) are calculated by the TOV equation. The training set of NN is filtered by astronomical observations. In the second step, the training set is substituted into the initialized NN training to examine the loss. Adjust the key parameters of NN until the loss converges. The final step is to find the minimum number of neurons at the input, which means comparing the relative errors under different mass-radius point conditions. The different mass-radius points come from the non-training sample, which is the EOS in the non-training set and the corresponding mass-radius (consistent with the parameter range and filtering in the first step).

The most significant prerequisite for the implementation of this algorithm is the training sample processing. Four groups of variable parameters are used to generate training samples to cover as large EOS range as possible. The variable parameter space is as follows: $30\leqslant L\leqslant 143\ \mathrm{MeV}$, $-400\leqslant K_\mathrm{sym}\leqslant 100\ \mathrm{MeV}$, $-200\leqslant J_\mathrm{sym}\leqslant 800\ \mathrm{MeV}$, $-300\leqslant J_0\leqslant 400\ \mathrm{MeV}$, $K_0=240\ \mathrm{MeV}$, $E_{\mathrm{sym}}\left( \rho _0 \right) = 31.7\ \mathrm{MeV}$ \cite{Shlomo33, Piekarewicz34, Zhang35, Oertel36, Li37}. It is especially worth noting that the slope $L$ range was developed by combining terrestrial \cite{Adhikari4} and astrophysical data. To obtain preliminary training samples, we generate EOS (Section \ref{ISOEOS}) by taking points at equal intervals in the above parameter interval, then solve the corresponding M-R with the TOV equation. Moreover, in order to make the output more reasonable, we used astronomical observation results during the sample generation stage: $M_{\max}\geqslant 2.14\ M_{\odot}$ \cite{Cromartie20NatAs}, $\varLambda _{1.4}=190_{-120}^{+390}$ \cite{Abbott18PRL}, $1.44\pm 0.15\ M_{\odot}$ \cite{Miller5} corresponding to $\left[ 11.2,\ 13.3 \right]\ \mathrm{km}$, $2.08\pm 0.07\ M_{\odot}$ \cite{Riley6} corresponding to  $\left[ 12.2,\ 16.3 \right]\ \mathrm{km}$. Similarly to Ref. \cite{Fujimoto_JHEP2021} considering the causality condition ($c_s<c$) and the stability condition ($dp/d\rho >0$), if the maximum mass filter is then added, the sample size is comparable to that after the screening of astronomical observations as described above, with a limited improvement in the generalization ability. In the sample processing stage, due to the randomness of observations, we randomly sampled the mass-radius sequences over the whole mass range. Note that because it is more difficult to form low-mass neutron stars, as in Ref. \cite{Fujimoto_JHEP2021} the sampling point is set to be greater than 1 $M_{\odot}$. But, considering the theoretical lower mass limit of neutron stars that can reach 0.1 $M_{\odot}$ \cite{Haensel_AA2002}, as well as recent observations of the HESS J1731-347 \cite{Doroshenko_NA2022}, a wider mass sampling interval was adopted. The final sample size involved in the NN is approximately 880,000, where M-R points are used as input and P-$\rho$ points are used as output. 

\begin{table}[H]
\caption{\textbf{Architecture of neural network}} \label{t}
\centering
\begin{tabular}{lcc}
\toprule
Layer &~~~~~~~~~Number of neurons&~~~~~~~~~Activation function \\
\midrule
0(Input)&~~~~~~~~~80 &~~~~~~~~~N/A \\
1&~~~~~~~~~100&~~~~~~~~~ReLU \\
2&~~~~~~~~~300&~~~~~~~~~ReLU \\
3&~~~~~~~~~400&~~~~~~~~~ReLU \\
4&~~~~~~~~~400&~~~~~~~~~ReLU \\
5&~~~~~~~~~400&~~~~~~~~~ReLU \\
6&~~~~~~~~~300&~~~~~~~~~ReLU \\
7(Output)&~~~~~~~~~182&~~~~~~~~~tanh \\
\bottomrule
\end{tabular}
\end{table}

The NN was initialized with reference to the framework of Ref. \cite{Fujimoto16} and \cite{Krastev15} and according to the background of our training set. By adjusting the relevant parameters, such as the number of layers, the optimization function, the epoch size, etc., the loss convergence of the NN is achieved. We finalized the NN architecture as shown in Table \ref{t}. The 80 neurons in the input layer means that the NN can provide accurate predictions under the condition of 40 pairs of (M,R) points. The other key parameters for this NN are shown in Appendix \ref{A} and the predicted results for the different input points are shown in Appendix \ref{B}.

To verify the accuracy, we used several mass-radius sequences of the EOS to substitute into this NN for prediction. In particular, we use APR3 and APR4 to test the effectiveness of this NN for non-parameter EOS.

\begin{figure}[htbp]
    \centering
    \includegraphics[scale=0.22]{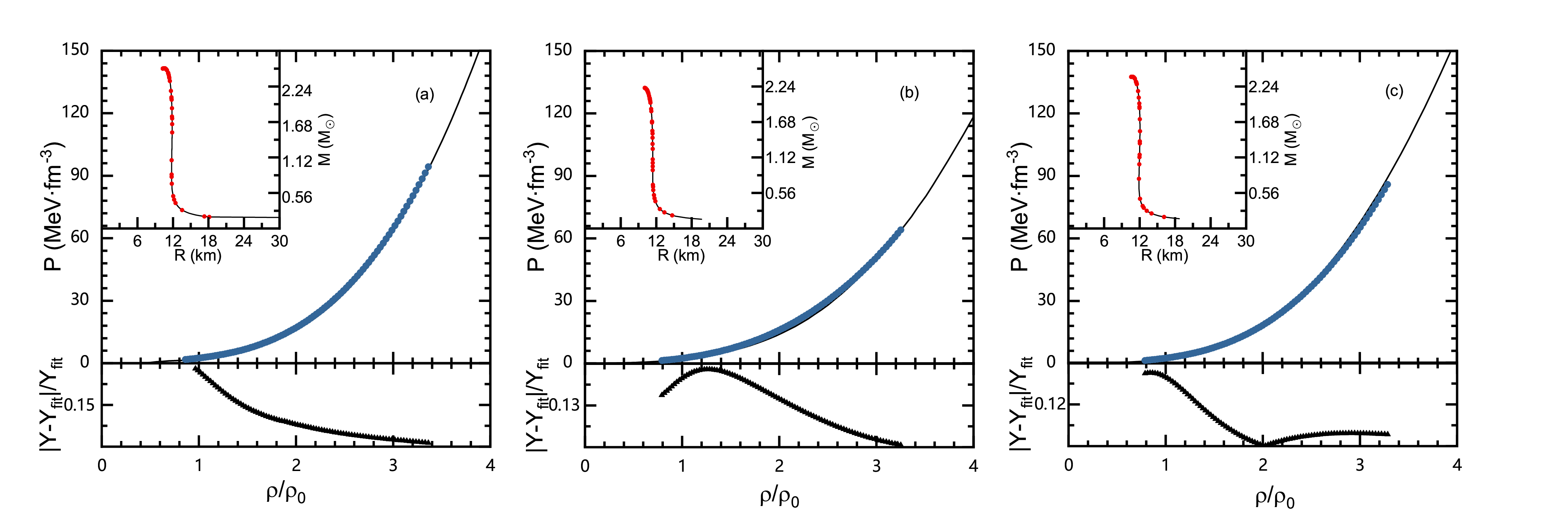}
    \caption{The EOS predictions of NN from 40 sampled points for the $L-J_0-J_{\mathrm{sym}}-K_{\mathrm{sym}}$ panel, where the solid black line represents the original EOS, and the blue dots denote the NN predictions. The three sample EOSs are, (a) parametric EOS with randomly generated parameters as $L = 40.5\ \mathrm{MeV}$, $K_{\mathrm{sym}} = -200.5\ \mathrm{MeV}$, $J_{\mathrm{sym}} = 320\ \mathrm{MeV}$, $J_0 = 110\ \mathrm{MeV}$, $K_0 = 240\ \mathrm{MeV}$, $E_{\mathrm{sym}}\left( \rho _0 \right) = 31.7\ \mathrm{MeV}$,  (b) APR3 EOS and (c) APR4 EOS. The inset shows the NN sampling from the mass-radius sequence. The lower panels are the relative error of the fitting.}
    \label{1}
\end{figure}

Fig. \ref{1} shows the NN predictions for three adopted EOS. It is shown that our NN method achieves a relatively accurate prediction function with approximately 40 mass-radius points as input. The solid black line for the (a) panel equation of state is randomly generated and substituted into Eq.(\ref{2.4}) to give the solid black line for the mass-radius sequence. Additionally, We also test some EOS other than the parameterized EOS and find that some of them can also be predicted well, as shown in (b) and (c) in Fig. \ref{1} for APR3 and APR4 EOS. This result means that our NN method will no longer be limited to parametric EOS even though we use only parametric EOS as training samples.


\begin{figure}[H]
    \centering
    \includegraphics[scale=0.42]{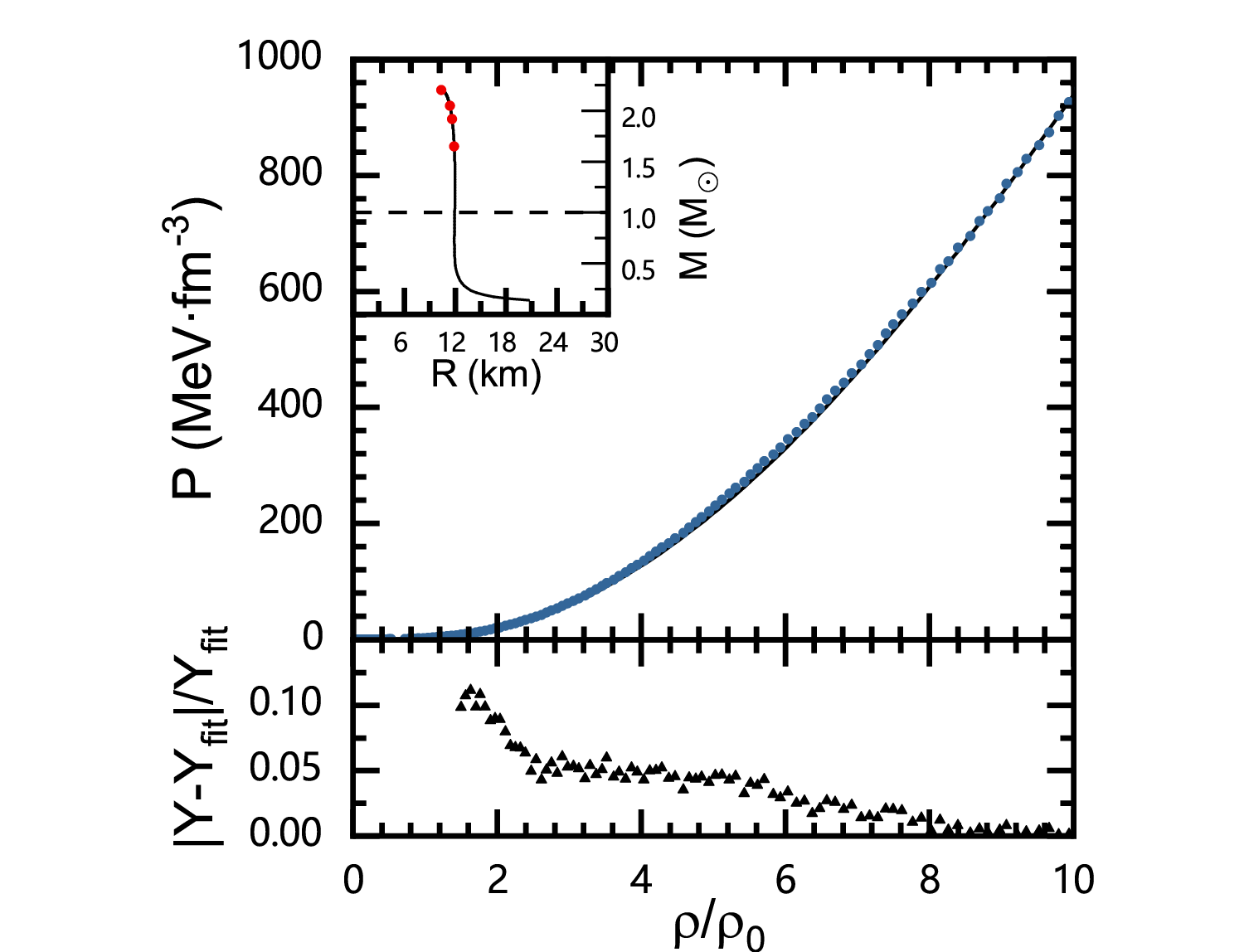}
    \caption{The EOS predictions of NN from 4 sampled points for the $L-K_{\mathrm{sym}}$ panel (The EOS parameters are as follows: $L = 40.5\ \mathrm{MeV}$, $K_{\mathrm{sym}} = 80.5\ \mathrm{MeV}$, $J_{\mathrm{sym}} = 0\ \mathrm{MeV}$, $J_0 = 0\ \mathrm{MeV}$, $K_0 = 240\ \mathrm{MeV}$, $E_{\mathrm{sym}}\left( \rho _0 \right) = 31.7\ \mathrm{MeV}$). The solid black line is the original EOS from which the sample points shown as the red dots were generated. The blue dots are the EOS data output by NN. The inset shows the NN sampling from the mass-radius sequence (Greater than 1\ $M_\odot$).}
    \label{2}
\end{figure}

For comparison with the results in Ref. \cite{Krastev15}, we also probe the output of NN by setting the parameters of EOS in the $L-K_{\mathrm{sym}}$ panel, where $J_0 = J_{\mathrm{sym}} = 0$, and  $L$ and $K_{\mathrm{sym}}$ take a range of $\left[ 30,\ 90 \right]\ \mathrm{MeV}$ and $\left[ -400,\ 100 \right]\ \mathrm{MeV}$, respectively.
Using the same sample generation process as for the $L-J_0-J_{\mathrm{sym}}-K_{\mathrm{sym}}$ panel, the sample size at this point is approximately 8,000. Fig. \ref{2} shows the NN predictions for the randomly selected test EOS with the parameters limited in the previously mentioned range.
We try to use as few input samples as possible to achieve the prediction function. Here the prediction function for the full density segment is implemented in the NN with 4 rows of mass-radius sequences as input (the data of the input is shown in Table \ref{t1}). Since the low-dimensional parameter space is simpler compared to the high-dimensional, e.g., the same (M, R) point corresponds to fewer EOS, only four output points and some parameters of the NN in the simplified Table \ref{t} are needed to implement the function. The above results show that our NN design is more suitable for inverse TOV mapping under isospin-dependent parametric EOS conditions.

\begin{table}[H]
\caption{\textbf{Mass-radius sampling points for Fig. \ref{2}}} \label{t1}
\centering
\begin{tabular}{cc}
\toprule
Mass ($M_\odot$)&~~~~~~~~~Radius (km) \\
\midrule
1.650030&~~~~~~~~~11.8838 \\
1.916370&~~~~~~~~~11.6289 \\
2.046519&~~~~~~~~~11.3741 \\
2.201511&~~~~~~~~~10.3368 \\
\bottomrule
\end{tabular}
\end{table}

Compared to Fig. \ref{1}, Fig. \ref{2} demonstrates a greater accuracy of the neural network predictions after reducing the two parameter dimensions. The fact that other different kinds of the EOS can be valid suggests that this NN prediction has certain universality.

\section{Combining NICER observations to verify the validity of NN} \label{s3}

\begin{figure}[H]
    \centering
    \includegraphics[scale=0.3]{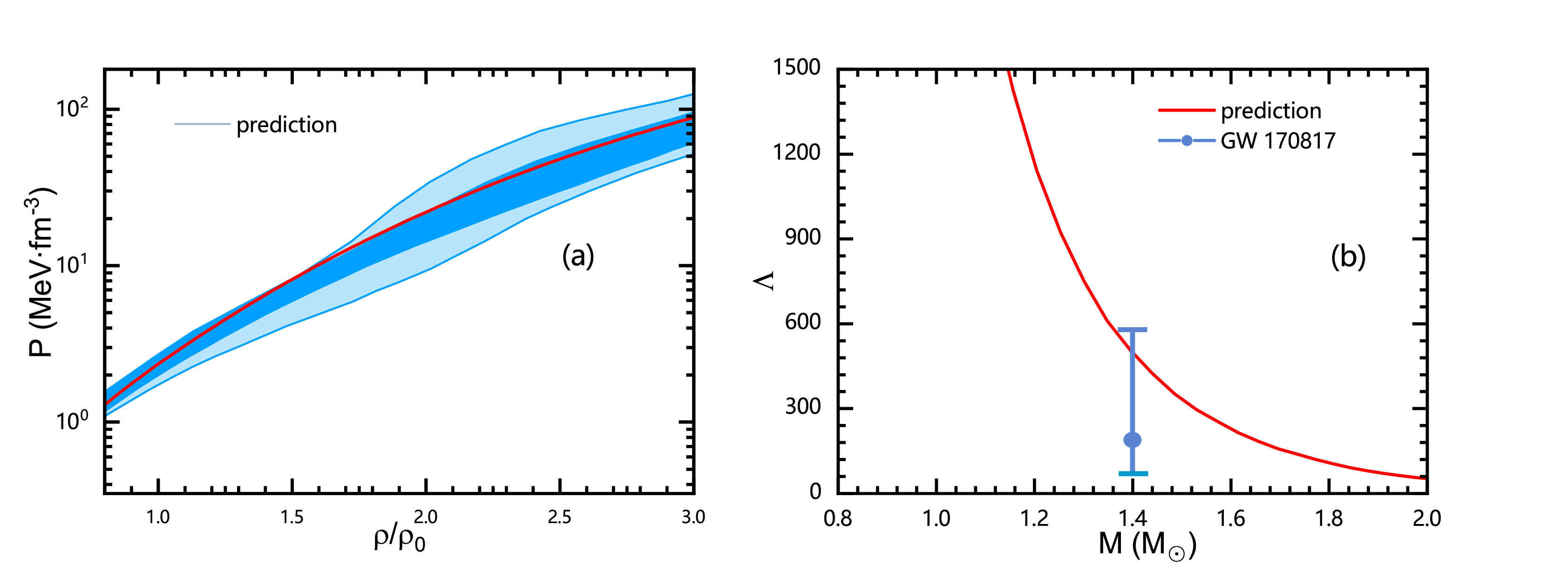}
    \caption{NN predictions combining the most probable points of the two NICER observations. (a) The red curve is NN's predicted EOS result, and the blue area presents the constraint of EOS from  Ref. \cite{Huth11} in  95\% (light) and 68\% (dark) credible intervals. (b) The red curve represents the dimensionless tidal deformation and stellar mass calculated using the EOS from (a). The constraint on the $\Lambda$ of canonical neutron star from  GW170817 \cite{Abbott8} is also presented.} 
    \label{3}
\end{figure}
To validate the predictions of the inverse TOV mapping-type NNs, prediction EOS from NICER observations are shown and discussed in this section within the parametric EOS framework as well as adopting the results of Refs. \cite{Abbott8} and \cite{Huth11}. Furthermore, due to the insufficient number of observations at the neutron star mass-radius, the following approach was adopted to satisfy the NN input points: 1. The EOS is generated in batch with parametric EOS within the range of nuclear parameters ($L$, $K_s$, $J_s$, $J_0$) mentioned in Sec. \ref{s2}; 2. Take one EOS that fits near the most probable point of the two NICER observations ($R_{1.44M\odot}=13.0\ km$ \cite{Miller21APJ}, $R_{2.08M\odot}=13.7\ km$ \cite{Riley19APJ}); 3. Forty points were randomly taken on the M-R curve corresponding to this EOS. However, for the real situation, 40 precise (M,R) observations from different neutron stars are necessary for NN input. It is expected that such accurate and sufficient observations will be available in the near future.

The NN-based predicted EOS together with further calculated dimensionless tidal deformations as a function of stellar mass are displayed in Fig. \ref{3}(a). To demonstrate the effect of NN, we also plot the constraint on EOS from Huth $et~ al.$ \cite{Huth11} and $\Lambda$ of canonical neutron star ($M=1.4M_{\odot}$) from GW170817 \cite{Abbott8}. It is shown that the predictions obtained under the assumption of observational accuracy prefer relatively stiff EOS at high density, which is basically in agreement with the results of Ref. \cite{Huth11}. Further calculation of the predicted EOS shows that the canonical neutron star has a dimensionless tidal deformation of $\varLambda _{1.4}\approx 507$, indicating a high tidal deformability. From Fig. \ref{3}(b), we can see that when the stellar mass  is less than 1.6 $M_{\odot}$, the tidal deformability decreases relatively quickly as the mass increases, when the mass is higher than 1.6 $M_{\odot}$, the tidal deformability decreases relatively slowly with the increase of mass. For massive neutron stars with $M=2.0M_{\odot}$, its tidal deformation is $\varLambda _{2.0}\approx 51$, which is far smaller than that of canonical neutron stars.

\section{Summary and Outlook} \label{s4}

The NN algorithms to implement the inverse mapping TOV equation is constructed by employing the parametric EOS as the training sample. Through the practical application of the obtained neural network algorithm, we get the following results.

\begin{itemize}
\item[(\romannumeral1)]
We implemented the inverse constraint EOS of NN in 4-dimensional parameter space ($L-J_0-J_{\mathrm{sym}}-K_{\mathrm{sym}}$) and 2-dimensional parameter space ($L-K_{\mathrm{sym}}$), respectively. About 40 mass-radius points are required to output higher precision EOS under the condition that $L-J_0-J_{\mathrm{sym}}-K_{\mathrm{sym}}$ is a variable. Four mass-radius points are required to output a higher precision EOS with $L-K_{\mathrm{sym}}$ as the variable.
\end{itemize}

\begin{itemize}
\item[(\romannumeral2)]
  The EOS based on the obtained NN algorithm and the two sets of  NICER  observational values with assumed precision is predicted. Similar to the results from other method, the EOS constrained by the NICER observation and NN is relatively stiff, and the corresponding tidal deformability of canonical neutron star is relatively high. Our NN predictions are basically consistent with the results of Huth $et~ al.$ \cite{Huth11}.

\end{itemize}

Despite the generalisation ability of the algorithm we provide, the output of the neural network is difficult to ensure credible results when the EOS is other than the isospin-dependent parametric EOS. To fill this gap,  more types of EOSs need  to be added to the training samples in the future. We believe that by combining precise observations from multiple sources, NN will be a promising tool for achieving more precise constraints on the EOS of nuclear matter.

\begin{acknowledgements}
This work is supported by NSFC (Grants No. 12375144 and 11975101), Guangdong Natural Science Foundation (Grants No. 2022A1515011552 and 2020A151501820).
\end{acknowledgements}

\appendix
\section{Neural network implementation}\label{A}
NN, one of the common algorithms in ML, is now used in almost every aspects of scientific research and engineering \cite{Wright28}. Based on the universal approximation theorem, NN can implement complex nonlinear mappings \cite{Hornik29, Leshno30}. Its construction is mainly divided into an input layer, a hidden layer and an output layer. Each neuron can be treated as a container of numbers. Similar to the transmission of electrical signals between synapses, neurons are using numbers as signals between them. Starting from the input layer, each layer of neurons goes through the following process between each other
\begin{equation}
z^\mathrm{l}=w_1a_{2}^{\mathrm{l-1}}+\cdots +w_\mathrm{n}a_{\mathrm{n}}^{\mathrm{l-1}}+b.
\end{equation}
where $w$ represents the weight, $l$ represents the serial number of layers, and $b$ represents the bias value \cite{Goodfellow}. The calculated $z$ is then substituted into the activation function to obtain the value of the output of the neuron to the next layer:
\begin{equation}
a_{\mathrm{j}}^{\mathrm{l}}=f\left( z^\mathrm{l} \right) ,
\end{equation}
where $f(x)$ is called the activation function and is usually used as Sigmoid or ReLU, the latter being adopted in this work. When summarizing the operations of the whole layer of neurons, we can obtain the following expression
\begin{equation}
f\left( \left[ \begin{matrix}
	w_{0,0}&		w_{0,1}&		\cdots&		w_{0,\mathrm{n}}\\
	w_{1,0}&		w_{1,1}&		\cdots&		w_{1,\mathrm{n}}\\
	\vdots&		\vdots&		\ddots&		\vdots\\
	w_{\mathrm{k},0}&		w_{\mathrm{k},1}&		\cdots&		w_{\mathrm{k},\mathrm{n}}\\
\end{matrix} \right] \left[ \begin{array}{c}
	a_{0}^{0}\\
	a_{1}^{0}\\
	\vdots\\
	a_{\mathrm{n}}^{0}\\
\end{array} \right] +\left[ \begin{array}{c}
	b_0\\
	b_1\\
	\vdots\\
	b_\mathrm{n}\\
\end{array} \right] \right) =output.
\end{equation}

At this point, the NN has completed one forward propagation. We also need a metric to evaluate how well the output compares to the true value. This type of evaluation metric is called the loss function, in which we usually use the mean squared error ($MSE=\frac{1}{\mathrm{n}}\sum_{\mathrm{i}=1}^\mathrm{n}{\left( Y_\mathrm{i}-\widehat{Y_\mathrm{i}} \right) ^2}$) method . The MSE of each batch is equivalent to the construction drawing, guiding the optimizer to continuously adjust a huge number of weights ($w$) and biases ($b$) in the NN. The available optimizers are Stochastic Gradient Descent (SGD) and Adam \cite{Adam}, the latter being adopted in this paper.

The platform for our implementation of NN is Keras, with Tensorflow as its backend \cite{Keras, Tensorflow}. It integrates the excellent features of Compute Unified Device Architecture (CUDA) for parallel computing on GPU with Tensorflow as the backend thus providing a rich interface. The training of the NN was performed on NVIDIA GeForce GTX 1650. It can significantly save the time cost of computing while implementing complex NN. To prevent overfitting, Dropout is set to 0.3 between layers. Furthermore, we chose an initial learning rate of 0.0003, a batch size of 500, and an epoch of 800. The proportion of the training dataset used for the validation dataset was 0.1. The hidden layer is used as a fully connected layer. Based on the above conditions, we have designed six layers of NN to achieve inverse TOV mapping. All the data in the NN need to be normalized, which in this paper is used as ({M}$/${3}, {R}$/${35}) and (($\log _{10}p$)$/${40}, $(\log _{10}\rho )/{20}$). The EOS is taken logarithmic in order to avoid too large a gap in pressure magnitude between different density intervals affecting the prediction accuracy of the NN. For ease of calculation both P and $\rho$ are taken as logarithmic results in $MeV\cdot fm^{-3}$ and $kg\cdot m^{-3}$ respectively.
\section{Results for different input points}\label{B}

\begin{figure}[H]
    \centering
    \includegraphics[scale=0.35]{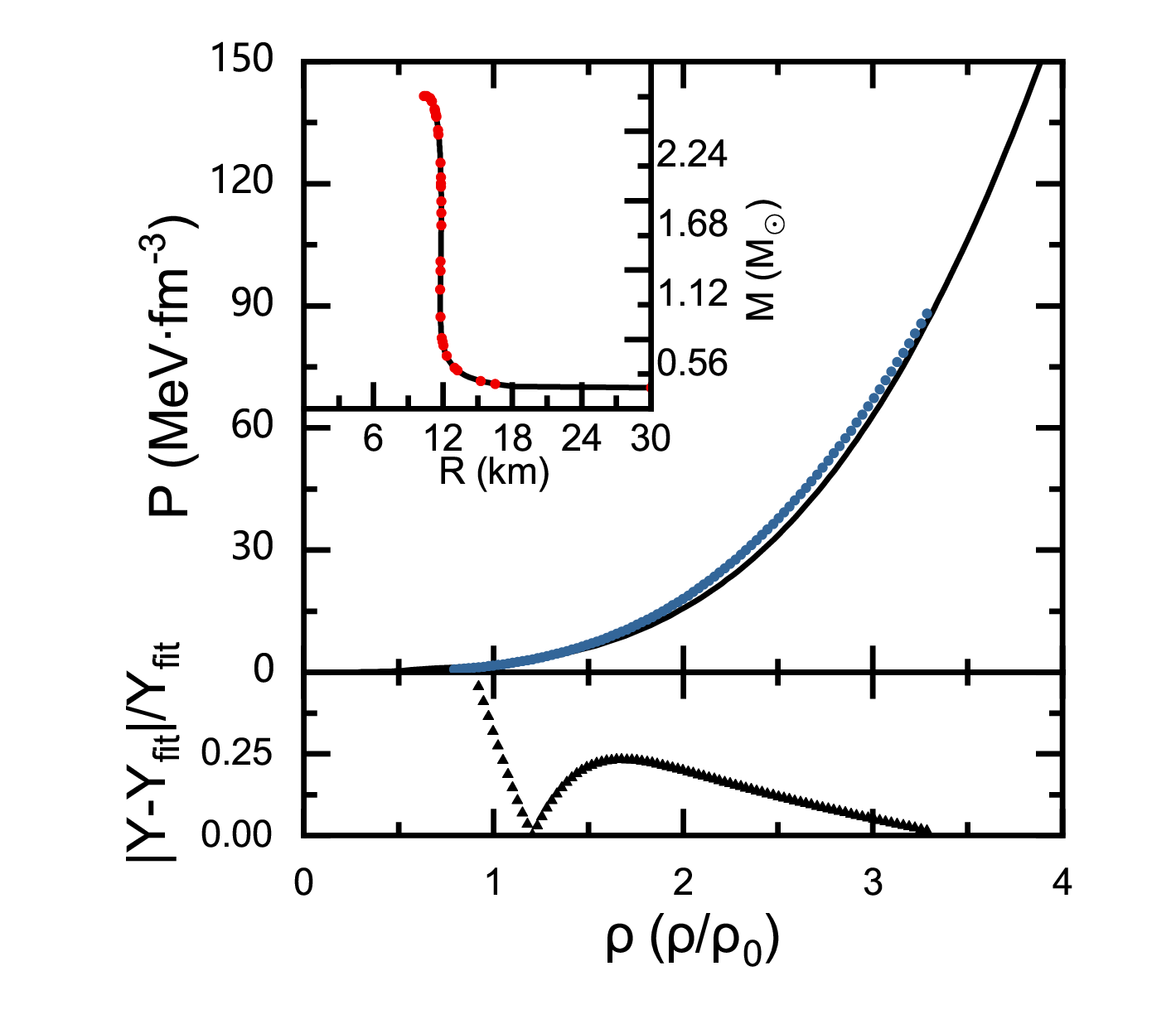}
    \caption{Similar to Fig \ref{1} (a), but for 35 sampled points.}
    \label{5}
\end{figure}

\begin{figure}[H]
    \centering
    \includegraphics[scale=0.35]{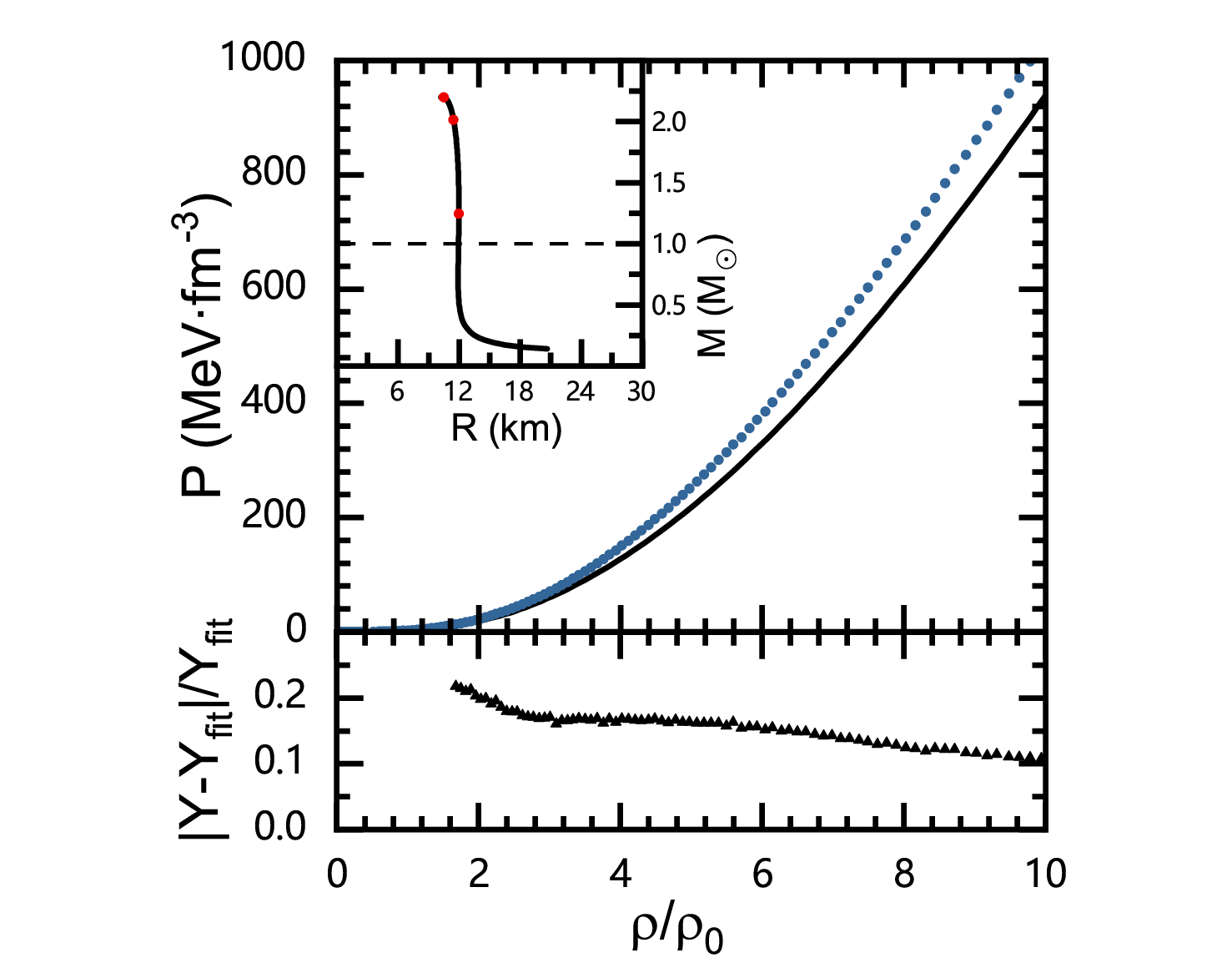}
    \caption{Similar to Fig. \ref{2}, but for 3 sampled points.}
    \label{6}
\end{figure}

We gradually reduce the input points after achieving loss convergence, while improving the relevant parameters of NN. Within the conventional NN framework, most of the improvements to achieve convergence of the NN can only affect the convergence rate of the loss cannot affect the minimum number of input points, such as increasing the number of hidden layers. As the number of neurons in the input layer continues to decrease, a significant error occurs in the $L-J_0-J_{\mathrm{sym}}-K_{\mathrm{sym}}$ panel for 35 pairs of (M,R) (see Fig. \ref{5}), while a significant error occurs in the $L-K_{\mathrm{sym}}$ panel for 3 pairs of (M,R) (see Fig. \ref{6}). It is important to note that over-concentration of (M,R) sampling points affects the NN predictions. Such sampling is a small probability event in astronomical observations.

$\\ \hspace*{\fill} \\$
\noindent
\textbf{Data Availability Statement:} The main code used in this work can be found in the Github repository \href{https://github.com/zhscut/NN-for-Neutron-Star}{https://github.com/zhscut/NN-for-Neutron-Star.}

\end{document}